\documentstyle[12pt]{article}
\newcommand{\be}{\begin{equation}}
\newcommand{\ee}{\end{equation}}
\newcommand{\ba}{\begin{eqnarray}}
\newcommand{\ea}{\end{eqnarray}}
\newcommand{\baa}{\begin{eqnarray*}}
\newcommand{\eaa}{\end{eqnarray*}}

\begin{document}
\thispagestyle{empty}

\vskip 2.0cm
{\renewcommand{\thefootnote}{\fnsymbol{footnote}}
\centerline{\large \bf Characteristic Temperatures of Folding of 
a Small Peptide}

\vskip 2.0cm
 
\centerline{Ulrich H.E.~Hansmann
\footnote{\ \ e-mail: hansmann@ims.ac.jp},
Masato Masuya
\footnote{\ \ e-mail: masatom@ims.ac.jp},
 and Yuko Okamoto
\footnote{\ \ e-mail: okamotoy@ims.ac.jp}} 
\vskip 1.5cm
\centerline {{\it Department of Theoretical Studies}} 
\centerline{{\it Institute for Molecular Science}} 
\centerline {{\it Okazaki, Aichi 444, Japan}}

\medbreak
\vskip 3.5cm
 
\centerline{\bf ABSTRACT}
\vskip 0.3cm
 We perform a {\it generalized-ensemble} simulation of a small peptide
taking the interactions among all atoms into account. From this
simulation we obtain thermodynamic quantities over a wide range of 
temperatures.  In particular,  
we show that the folding  of a small peptide is a multi-stage
process  associated with two characteristic temperatures, the collapse
temperature $T_{\theta}$ and the folding temperature $T_f$. 
Our results  give supporting evidence for the energy landscape picture
and funnel concept. These ideas were previously 
developed in the context of studies 
of simplified protein models, and here 
for the first time
checked in an all-atom Monte Carlo simulation.
\vfill
\newpage}
 \baselineskip=0.8cm
\noindent

It is well known that a large class of proteins fold spontaneously 
into globular states of unique shape, yet the mechanism of  folding
has remained elusive.  The folding process may be either thermodynamically
or kinetically controlled. The ``thermodynamic hypothesis''
 assumes that the folded state corresponds to the global minimum in free
energy and is supported by  the famous work of Anfinsen \cite{Anf} and similar
experiments. On the other hand, Levinthal \cite{Lev} has argued that 
because of the 
huge number of local energy minima available to a protein, it is
impossible to find the global free energy minimum  by a random search
in biological time scales (of order seconds). His argument rather
suggests that the protein folds into a unique metastable state, the
kinetically most accessible structure. 
The complexity and importance of the problem raised a lot of interest in the 
subject over the last three decades, but no complete solution
is in sight to date. However, significant new insight was gained over the
last few years from the studies of minimal protein models. 
Both lattice models \cite{Go}--\cite{SKO2} and continuum models 
\cite{LW}--\cite{Fuk} have been extensively studied.
Common to all these models is that they   capture
only few, but probably dominant interactions in real proteins. 
These include chain connectivity, excluded volume,
 hydrophobicity as the driving force,  and
sequence heterogeneity. For recent reviews on minimal protein models and
their applications, see Refs.~\cite{BOSW}--\cite{SHA4}. From the numerical and 
analytical studies of these models  a new view  of the folding process 
emerged.  The folding kinetics is seen to be determined by an energy 
landscape which for foldable proteins 
resembles a funnel with a free energy gradient toward the native 
% NEW
structure \cite{LMO,OWLS,SOW,BOSW,DC}. 
The funnel is itself  rough and folding occurs by a multi-stage, multi-pathway
kinetics. 
%kinetics \cite{LMO,OWLS,SOW,BOSW,DC}. 
% END NEW
A common scenario for folding may be that first the polypeptide 
chain collapses from a
random coil to a compact state. This coil-to-globular transition is 
characterized by the
collapse transition temperature $T_{\theta}$. In the second stage,
a set of compact structures is explored. The final stage involves a
transition from one of the many local minima in 
the set of compact structures to the native conformation. This final transition
is characterized by the folding temperature $T_f~ (\le T_{\theta})$. 
It was conjectured
that  the kinetic accessibility of the native conformation can be classified
by the parameter \cite{THI4} 
\begin{equation}
%\sigma = (T_{\theta} - T_f)/T_{\theta};
% NEW
\sigma = \frac{T_{\theta} - T_f}{T_{\theta}}~,
% END NEW
\label{sig}
\end{equation}
i.e., the smaller $\sigma$ is,
the more easily a protein can fold. 
If $T_{\theta} \approx T_f$ (i.e., $\sigma \approx 0$), the second stage 
will be very short or not exist, and the protein will fold in an ``all or 
nothing'' transition from the extended coil to the native conformation without
any detectable intermediates. On the other hand, for some proteins the
folding process may involve further stages.  A more elaborate classification
of  possible folding processes is discussed in Ref.~\cite{BOSW}.

One can ask whether the picture outlined above is really useful to
describe the folding kinetics of real proteins, because the underlying
models are gross simplifications of real protein systems. For instance, 
side-chain conformational degrees of freedom that are 
important for packing are
neglected. The situation actually resembles a vicious circle. The energy
landscape picture and the analogy to phase transitions were developed from 
studies of the highly simplified description of proteins by minimal models. 
However, only if these concepts apply for proteins, it is possible to argue
that the broad mechanism of phase  transitions depends solely on gross features
of the energy function, not on their details. 
Only in this case  a law of corresponding states 
can be applied to explain dynamics of real proteins from studies of the
folding kinetics in minimal models. It is therefore
desirable to check  the above picture by comparison with 
more realistic energy functions, namely, with
all-atom simulations of a suitable protein. This is the purpose of 
the present article.
% NEW
While there has been an attempt to study the
free energy landscape of an all-atom protein model 
 by unfolding MD simulations \cite{Brooks},  the present work starts
from random initial conformations and is rather concerned with obtaining
%Free energy landscape of an all-atom protein model has been directly
%studied by unfolding MD simulations \cite{Brooks}.
%The present work is rather concerned with obtaining the two
characteristic temperatures of protein folding by a Monte Carlo
simulation 
%simulation from a random initial conformation 
(and thus studying the energy landscape indirectly).
% END NEW

Simulations of proteins where the interactions among all atoms
are taken into account have been notoriously difficult (for a recent review, 
see Ref.~\cite{Vas}). 
The various competing 
interactions yield to a much rougher energy landscape than for minimal 
protein models. 
 (In fact, one might question whether the limitations of the 
% NEW
%current energy functions lead not to much rougher energy landscapes than
 current energy functions may lead to rougher energy landscapes than
% END NEW
 the protein encounters in nature.) 
Simulations based on canonical 
Monte Carlo or molecular dynamics techniques will at low
temperatures get trapped in one of the  multitude of local minima separated
by high energy barriers. Hence, only small parts of configuration space 
are sampled  
and physical quantities cannot be calculated accurately. However, with the
development of  {\it generalized-ensemble} techniques
like multicanonical
algorithms \cite{MU} and simulated tempering \cite{ST,ST2},  
an efficient sampling of low-energy configurations and calculation of 
accurate low-temperature thermodynamic  quantities  became feasible. 
The first application of one of these
techniques to the protein folding problem can be found in Ref.~\cite{HO}. 
Later applications of multicanonical algorithms 
include the study of the coil-globular
transitions of a simplified model protein \cite{HSC} and the 
helix-coil transitions of
homo-oligomers of nonpolar amino acids 
\cite{HO95a}. 
A formulation of multicanonical algorithm 
for the molecular dynamics method
was also developed \cite{HO96c,NNK}.
A numerical comparison
of three different generalized-ensemble algorithms can be found in 
Ref.~\cite{HO96b}.  

The generalized-ensemble technique we utilize in this article was first 
introduced in Refs.~\cite{H97a,HO96d} and is related to Tsallis generalized 
mechanics formalism \cite{Tsa}. In this algorithm, configurations are
updated according to the following probability weight:
\begin{equation}
w(E) = \left(1+ \frac{\beta (E-E_0)}{n_F}\right)^{-n_F}~,
\label{eqwe}
\end{equation}
where $E_0$ is an estimator for the ground-state energy, $n_F$ is 
the number of degrees of freedom of the system, and $\beta = 1/k_BT$ 
is the inverse temperature with a low temperature $T$ (and $k_B$ is the 
Boltzmann constant).
Obviously, the new weight reduces in the low-energy  
region to the canonical Boltzmann weight $\exp (- \beta E)$ for 
$\frac{\beta (E-E_{0})}{n_F} \ll 1$. 
 On the other hand, high-energy regions are no
longer exponentially suppressed but only according to a power law,
which enhances excursions to high-energy regions. 
In contrast to other  generalized-ensemble techniques 
where the determination of weights is non-trivial,
the weight of the new ensemble  is explicitly given by Eq.~(\ref{eqwe}).
One only needs to find  an estimator for the ground-state energy $E_0$ 
which can be done by a procedure described in Ref.~\cite{HO96d} and is
much easier than the determination of weights for other generalized
ensembles.
Since the simulation by the present algorithm samples a
large range of energies, we can
use the reweighting techniques \cite{FS} to construct
canonical distributions and calculate thermodynamic 
average of any physical quantity $\cal{A}$
over a wide temperature range:
\begin{equation}
<{\cal{A}}>_T ~=~ \frac{\displaystyle{\int dx~{\cal{A}}(x)~w^{-1}(E(x))~ 
                 e^{-\beta E(x)}}}
              {\displaystyle{\int dx~w^{-1}(E(x))~e^{-\beta E(x)}}}~,
\label{eqrw}
\end{equation}
where $x$ stands for configurations.

Here, we use these novel techniques to examine the picture
for the folding kinetics as proposed from the simulations of minimal
models. Limitations in available computational time force us to restrict
ourselves on the simulation of small molecules, and we have in addition
 neglected explicit solvent interactions. The system of our choice is 
Met-enkephalin, one of the simplest peptides, with which we are 
familiar from earlier work \cite{HO,HO96b,HO94_3}.  
Met-enkephalin has the amino-acid sequence Tyr-Gly-Gly-Phe-Met.
The potential energy function
$E_{tot}$ (in kcal/mol) that we used is given by the sum of
the electrostatic term $E_{es}$, 12-6 Lennard-Jones term $E_{LJ}$, and
hydrogen-bond term $E_{hb}$ for all pairs of atoms in the peptide together 
with the torsion term $E_{tors}$ for all torsion angles:
\begin{eqnarray}
E_{tot} & = & E_{es} + E_{LJ} + E_{hb} + E_{tors},\\
E_{es}  & = & \sum_{(i,j)} \frac{332q_i q_j}{\epsilon r_{ij}},\\
E_{LJ} & = & \sum_{(i,j)} \left( \frac{A_{ij}}{r^{12}_{ij}}
                                - \frac{B_{ij}}{r^6_{ij}} \right),\\
E_{hb}  & = & \sum_{(i,j)} \left( \frac{C_{ij}}{r^{12}_{ij}}
                                - \frac{D_{ij}}{r^{10}_{ij}} \right),\\
E_{tors}& = & \sum_l U_l \left( 1 \pm \cos (n_l \chi_l ) \right),
\end{eqnarray}
where $r_{ij}$ (in \AA) is the distance between the atoms $i$ and $j$,
and $\chi_l$ is the $l$-th torsion angle.
The parameters ($q_i,A_{ij},B_{ij},C_{ij},
D_{ij},U_l$ and $n_l$) for the energy function were adopted
from ECEPP/2 \cite{EC3}.
The dielectric constant $\epsilon$ was set equal to 2.
% NEW
In ECEPP/2 bond lengths and bond angles 
%(which are hard degrees of freedom) 
are fixed at experimental values. 
%and no out-of-plane deformation of peptide
%bonds is allowed leaving the dihedral angles $\phi,\psi,\omega $ and
%$\chi$ as independent variables.
We further fix the peptide bond angles $\omega$ 
%(which are also hard degrees of freedom) 
 to their common value $180^{\circ}$, which 
%(its common value for amino acids other than proline)
 leaves us with 19 torsion angles ($\phi,~\psi$, and $\chi$) as independent
degrees of freedom (i.e., $n_F = 19$).
% END NEW
The computer code KONF90 \cite{KONF}  was
used.  We remark that KONF90 uses a different convention
for the implementation of the
 ECEPP parameters (for example, $\phi_1$ of ECEPP/2 is equal to
 $\phi_1 - 180^{\circ}$ of KONF90).  Therefore, our energy values are
slightly different
from those of the original implementation
of ECEPP/2.
The simulation was started from a completely random initial
conformation (Hot Start).  One Monte Carlo sweep updates every torsion angle
of the peptide once.
   
It is known from our previous work that the ground-state conformation 
for Met-enkephalin has the
KONF90 energy value
$E_{GS} = -12.2$ kcal/mol \cite{HO94_3}. We therefore set $E_0 = -12.2$ 
kcal/mol 
 and $T = 50$ K (or, $\beta = 10.1$ $[\frac{1}{{\rm kcal}/{\rm mol}}]$)
(and $n_F =19$) in our probability weight factor in Eq.~(\ref{eqwe}).
The ground-state structure, exhibiting a II'-type $\beta$ turn,  
is shown in Fig.~1.  
It is a superposition of ball-and-stick and space-filling representations.
The latter representation was added in order to give a rough idea of
the  volume of the peptide as discussed below.
     
All thermodynamic quantities were then  calculated from
 a single production run of 1,000,000 MC sweeps
which followed 10,000 sweeps
for thermalization. At the end of every
fourth sweep we stored the 
energies of the conformation, the corresponding volume, and  
the overlap of the conformation with the (known) ground state
for further analyses. Here, we approximate the volume of the peptide 
by its solvent excluded volume  (in \AA$^3$) which is  calculated by a 
 variant \cite{MO} of the double cubic lattice method \cite{ELASS}.  
Our definition of the overlap, which measures how much a given conformation
differs from the ground state, is given by
\begin{equation}
O(t) = 1 -\frac{1}{90~n_F} \sum_{i=1}^{n_F} |\alpha_i^{(t)}- \alpha_i^{(GS)}|~,
\label{eqol}
\end{equation}
where $\alpha_i^{(t)}$ and $\alpha_i^{(GS)}$ (in degrees) stand for 
the $n_F$ dihedral angles of the conformation at $t$-th Monte Carlo sweep 
and the ground-state conformation, respectively. Symmetries
for the side-chain angles were taken into account and the difference
$\alpha_i^{(t)}- \alpha_i^{(GS)}$ was always projected into the interval
$[-180^{\circ},180^{\circ}]$. Our definition  guarantees that we have 
\begin{equation}
0 \le ~<O>_T~ \le 1~,
\end{equation}
with the limiting values
\begin{equation}
\left\{
\begin{array}{rl}
 <O(t)>_T~~ \rightarrow 1~,~~&T \rightarrow 0~, \\
 <O(t)>_T~~ \rightarrow 0~,~~&T \rightarrow \infty~.
\end{array}
\right.
\end{equation}
~\\

Let us now present our results. In Fig.~2a we show the \lq \lq time series"
 of the total potential energy $E_{tot}$.
 It is a random walk in potential energy
space, which keeps the simulation from getting trapped in a local minimum.
It indeed visits the low-energy region 
several times
in 1,000,000 Monte Carlo sweeps. The visits are separated by
excursions into high-energy regions, which ensures de-correlation of
the configurations. This can be seen in Figs.~2b and 2c,
 where time series of the excluded volume and
the overlap function are displayed. 
The large changes in these quantities 
imply the large conformational changes which happen in the course of
the simulation. Since large parts of the configuration space are sampled,
the use of the reweighting techniques \cite{FS} is justified to 
calculate thermodynamic quantities over a wide range of temperatures
by Eq.~(\ref{eqrw}).

We expect the folding of
proteins and peptides to occur in a multi-stage process. The first process 
is connected with a collapse of the extended coil structure into 
an ensemble of compact structures. This transition should be connected
with a pronounced change in the average potential energy as a function of
temperature. At the transition temperature we therefore expect 
 a peak in the specific heat. Both quantities are shown in Fig.~3.
We clearly 
observe a steep decrease in total potential
energy around $300$ K and the corresponding
peak in the specific heat defined by
\begin{equation}
  C \equiv \frac{1}{N~k_B} \frac{d \left( <E_{tot}>_T \right)}{d T}
= {\beta}^2 \ \frac{<E_{tot}^2>_T - <E_{tot}>_T^2}{N}~,
\label{eqsh}
\end{equation}
where $N \ (=5)$ is the
number of amino-acid residues in the peptide. 
In Fig.~4 we  display the average values 
of each of the component terms of the potential energy 
(defined in Eqs.~(5)--(8)) 
as a function of temperature. 
% NEW
As one can see in the Figure, the change in average 
% END NEW
potential energy is 
mainly caused by the Lennard-Jones term 
% NEW
%(while the other terms vary much less with temperature) 
% END NEW
and therefore is connected to a decrease of
the  volume occupied by the peptide. This can be seen in Fig.~5,
 where we display the average volume as a function of temperature.
As expected, the volume decreases rapidly 
in the same temperature range as the potential
energy.  The average  volume is a natural measure of compactness,  but the 
change from  extended coil structures to  compact structures with decreasing
temperature can also be observed in other quantities like the 
average end-to-end distance $<d_{e-e}>_T$
(here, defined to be
the distance between N of Tyr$^1$ and O of Met$^5$). 
In Table~I, we give some of the values of $<d_{e-e}>_T$ as a function
of temperature.  The results imply again that the peptide is quite
extended at high temperatures and compact at low temperatures.

If both energy and volume decrease are correlated, then the
transition temperature $T_{\theta}$ can be located both from the
position where the specific heat has its maximum and from the
position  of the maximum of
\begin{equation}
\frac{d<V>_T}{dT} \equiv 
\beta^2 \left( <V E_{tot}>_T - <V>_T <E_{tot}>_T \right)~,
\end{equation}
which is also displayed in Fig.~5. The second quantity measures the steepness
of the decrease in volume in the same way as the specific heat measures the
steepness of decrease of potential energy.  To quantify its value we divided
our time series in 4 bins corresponding to 250,000 sweeps each,
determined the  position of the maximum for both quantities 
in each bin and averaged over the bins. In this way we found a
transition temperature $T_{\theta} = 280 \pm 20$ K from the location of
the peak in specific heat and $T_{\theta} = 310 \pm 20$ K from the
maximum in $d<V>_T/dT$. Both temperatures are indeed consistent with each
other within the error bars. 

The second transition which should occur at a lower temperature $T_f$
is that from a set of compact structures to the ``native 
conformation'' that is considered to be the ground
state of the peptide. Since
these compact conformations are  expected  to be all of similar volume and
energy 
% NEW
(systematic comparisons of such structures were tried in
 previous work \cite{Braun,OK,EH96}),
% END NEW
we do not expect to see this transition by pronounced changes
in $<E_{tot}>_T$ or to find another peak in the specific heat. Instead this
transition should be characterized by a rapid change in the average overlap
$<O>_T$ with the ground-state conformation (see Eq.~(\ref{eqol})) 
and a corresponding maximum in 
\begin{equation}
\frac{d<O>_T}{dT} \equiv 
\beta^2 \left( <O E_{tot}>_T - <O>_T <E_{tot}>_T \right)~.
\end{equation}
Both quantities are displayed in Fig.~6, and
we indeed find the expected behavior. The change in the order parameter
is clearly visible and occurs at a temperature lower than the 
first transition temperature $T_{\theta}$. We again try to determine its value
by searching for the peak in $d<O>_T/dT$ in each of the 4 bins and averaging
over the obtained values. In this way we find a transition temperature
of $T_f = 230 \pm 30$~K. 
% NEW
%It is interesting to observe that  $d<O>_T/dT$ approaches its limiting
%value zero for $T \rightarrow 0$ very slowly. This is because even at
%our lowest temperatures ( $T=50$ K) the configurations are not frozen
%Small variations are possible without that the overall structure
%changes as was observed in Ref.~\cite{HO94_3}. The slow decay of the
%derivative may therefore be interpreted as an indication for glassy
%behavior.  While our data do not allow to proof such behavior or
%even determine the glass temperature, the low temperature behavior of
%$d<O>_T/dT$ shows that such a glass phase will exist only for $T < T_f$
% END NEW
We remark that the average overlap $<O>_T$
approaches its limiting
value zero only very slowly as the temperature increases. This is 
because $<O>_T ~= 0$ corresponds to
a completely random distribution of dihedral angles which is energetically 
highly unfavorable because of the steric hindrance of both main and side
chains.

One characterization of the folding properties
of a peptide or protein is given by the parameter $\sigma$ of Eq.~(\ref{sig}).
With our values for $T_{\theta}$ and $T_f$, we have for Met-enkephalin
$\sigma \approx 0.2$.  Here, we have taken the central values:
$T_{\theta} = 295$ K and $T_f = 230$ K. 
%NEW
%According to Ref.~\cite{THI2} and \cite{THI4}  
%END NEW
This value of $\sigma$ implies that our peptide has reasonably 
good folding properties
%NEW
according to Refs.~\cite{THI2} and \cite{THI4}.  
%END NEW
 We remark that the characterization of Met-enkephalin as a good folder 
 has to be taken with care:
 Low-temperature simulations of the 
 molecules with conventional methods are still a formidable task
%NEW
and  a low value of $\sigma$ may not neccessarily indicate easy
foldability in a computer simulation.
%END NEW

%It is interesting to observe that  
While the collapse temperature $T_{\theta}$
 is roughly equal to room temperature,  the transition temperature $T_f$ is
 well below room temperature. Consequently, contributions of 
 ground-state conformers are not  dominant at room temperature 
for Met-enkephalin, as was 
% NEW
 observed in our earlier work \cite{HO,HO94_3}. This is due to the
small size of the peptide. However,  it still can be regarded as
a good model for a small protein, since it has a unique stable structure
below $T_f$. It was shown in Refs.~\cite{HO,HO94_3} that 
%However, the ground-state structure is not negligible either at
%this temperature.
%The probability of this structure to be found
%at $T \approx 300$ K is as high as 20 to 30 \% \cite{HO,HO94_3}.  
%it was also shown that Met-enkephalin remains mainly in the
 Met-enkephalin remains mainly in the
 vicinity of the ground state without getting trapped in any of
 the local-minimum structures below $T_f~(\approx 230$ K).
%~(\approx 230$ K) \cite{HO,HO94_3}.
%These results again suggest that Met-enkephalin has good folding properties
%and that the glass transition temperature $T_G$ is 
%lower than $T_f$ (or, probably there is no glass phase for 
%this system).
% END NEW
 
We also performed a generalized-ensemble 
simulation with the same statistics 
for a second peptide, Leu-enkephalin (data not shown). 
We found: $T_{\theta} = 300 \pm 30$ K and $T_f = 220 \pm 30$ K.
These transition temperatures are essentially the 
same as for Met-enkephalin. Both peptides are very similar, differing only 
in the side chains of the Met (Leu) residue. Our results indicate that 
indeed the general mechanism of the transition does not depend on 
these details and a law of corresponding state can be applied for 
similar peptides.

Let us summarize our results. We have  performed a generalized-ensemble
simulation of a small peptide taking the interactions among all
atoms into account and calculated thermodynamic averages
of physical quantities
over a wide range of temperatures. 
We found for this  peptide  two characteristic temperatures. 
% NEW
%The higher one
The higher temperature
% END NEW
is associated with a collapse of the peptide from extended coils into more 
compact structures, whereas the second one  indicates the transition between
an ensemble of compact structures and a phase which is dominated by a
single conformation, the ground state of the peptide. Our results
support
pictures for the kinetics of protein folding which were developed from
the study, both numerical and analytical, of simplified  protein models.
It is still an open question whether these minimal models can be used for
predictions of protein conformations.  However, our analyses, 
performed with an energy
function which takes much more of the details of a protein into account,
demonstrate that these models are indeed able to describe the
general mechanism of the folding process. Hence, the study of simplified
models  may in this way guide 
further simulations with more realistic energy functions.
The present paper aims to be a first step in this direction.

\vspace{0.5cm}
\noindent
{\bf Acknowledgments}: \\
Our simulations were  performed on computers 
of the Institute for Molecular Science (IMS), Okazaki,
Japan.
This work is supported by Grants-in-Aid for Scientific Research from the
Japanese Ministry of Education, Science, Sports, and Culture.

%%%%%%%%%%%%%%%%%%%%%%%%% references %%%%%%%%%%%%%%%%%%%

\newpage
~~\\

\noindent
{\bf Table~I.}  Average end-to-end distance $<d_{e-e}>_T$ (\AA) of
Met-enkephalin as a function of temperature $T$ (K). \\
\vspace{-0.3cm}
\begin{center}
\begin{tabular}{cccccccccc} \hline
\hline
\rule{0mm}{3mm} \\
$T$ & 50 & 100 & 150 & 200 & 250 & 300 & 400 & 700 & 1000 \\ 
\rule{0mm}{3mm} \\
\hline
\rule{0mm}{3mm} \\
$<d_{e-e}>_T$ & 4.8 & 4.8 & 4.9 & 5.2 & 5.8 & 6.8 & 8.6 & 11.0 & 11.5 \\ 
\rule{0mm}{3mm} \\
\hline
\hline
\end{tabular}
\end{center}
\noindent
        
\newpage
\centerline{\bf Figure Legends}

\begin{itemize}
\item FIG.~1: Ground-state conformation of Met-enkephalin for
 KONF90 energy function.
The figure was created with
      Molscript \cite{Molscript} and Raster3D \cite{Raster3D}.
\item FIG.~2: Time series of total potential energy $E_{tot}$ (kcal/mol)
      (a), volume $V$ (\AA$^3$) (b), and overlap $O$
              (defined by Eq.~(\ref{eqol})) (c) as obtained by a 
               generalized-ensemble
               simulation of 1,000,000 Monte Carlo sweeps.
\item FIG.~3: Average total potential energy $<E_{tot}>_T$ 
              and specific heat C 
              as a function of temperature. 
              The dotted vertical line is added to aid the eyes 
              in locating the peak of specific heat.  
              The results were obtained from a
              generalized-ensemble simulation of 1,000,000 Monte Carlo sweeps.
 \item FIG.~4: Average potential energies  
               as a function of temperature. The results were obtained from a
               generalized-ensemble simulation of 1,000,000 Monte Carlo sweeps.
 \item FIG.~5: Average volume $<V>_T$  and its derivative $d<V>_T/dT$ 
               as a function of temperature. 
              The dotted vertical line is added to aid the eyes 
              in locating the peak of the derivative of volume.  
               The results were obtained from a
               generalized-ensemble simulation of 1,000,000 Monte Carlo sweeps.
\item FIG.~6: Average overlap $<O>_T$ and its derivative $d<O>_T/dT$ 
              as a function of temperature. 
              The dotted vertical line is added to aid the eyes 
              in locating the peak of the derivative of overlap.  
              The results were obtained from a
              generalized-ensemble simulation of 1,000,000 Monte Carlo sweeps.
\end{itemize}
%%%%%%%%%%%%%%%%%%%  

\noindent
\begin{thebibliography}{(00)}
\bibitem{Anf} Anfinsen, C. B. (1973) {\it Science} {\bf 181}, 223--230.
\bibitem{Lev} Levinthal, C.\ (1968) {\it J. Chem. Phys.} {\bf 65}, 44--45.
\bibitem{Go} Taketomi, H., Ueda, Y. \& G\={o}, N. (1975) {\it
Int. J. Peptide Protein Res.} {\bf 7}, 445--459.
\bibitem{SKO0} Skolnik, J. \& Kolinski, A. (1990) {\it Science} {\bf
250}, 1121--1125.
\bibitem{Dill0} Chan, H. S. \& Dill, K. A. (1991) {\it
Annu. Rev. Biophys. Biophys. Chem.} {\bf 20}, 447--490.
\bibitem{SHA1} Shakhnovitch, E. I., Farztdinov, G. M., Gutin, A. M. \&
Karplus, M. (1991) {\it Phys. Rev. Lett.} {\bf 67}, 1665--1668.
\bibitem{Dill1} Miller, R., Danko, C. A., Fasolka, M. J., Balazs, A. C.,
 Chan, H. S. \& Dill, K. A. (1992) {\it J. Chem. Phys.} {\bf 96},
 768--780.
\bibitem{LMO} Leopold, P. E., Montal, M. \& Onuchic, J. N. (1992) 
{\it Proc. Natl. Acad. Sci. USA} {\bf 89}, 8721--8725.
\bibitem{THI2} Camacho, C. J. \& Thirumalai, D. (1993) {\it
Proc. Natl. Acad. Sci. USA} {\bf 90}, 6369--6372.
%\bibitem{SKO1} Kolinski, A. \& Skolnik, J. (1994) {\it Proteins} {\bf
%18}, 338--352.
\bibitem{SHA3} Sali, A., Shakhnovitch, E. I. \& Karplus, M. (1994) {\it
 J. Mol. Biol.} {\bf 235}, 1614--1636.
\bibitem{HSC} Hao, M.-H. \& Scheraga, H. A. (1994) {\it J. Phys. Chem.} 
{\bf 98}, 4940--4948.
\bibitem{OWLS} Onuchic, J. N., Wolynes, P. G., Luthey-Schulten, Z. \&
Socci, N. D. (1995) {\it Proc. Natl. Acad. Sci. USA} {\bf 92},
3626--3630.
\bibitem{SOW} Socci, N. D., Onuchic, J. N. \& Wolynes, P. G. (1996) {\it 
 J. Chem. Phys.} {\bf 104}, 5860--5871.
\bibitem{THI4} Klimov, D. K. \& Thirumalai, D. (1996) {\it
Phys. Rev. Lett.} {\bf 76}, 4070--4073.
\bibitem{SKO2} Kolinski, A., Galazka, W. \& Skolnik, J.\ (1996) {\it
Proteins} {\bf 26}, 271--287.
\bibitem{LW} Levitt, M. \& Warshel, A. (1975) {\it Nature (London)} {\bf
253}, 694--698.
\bibitem{BW1} Bryngelson, J. D. \& Wolynes, P. G. (1987) {\it
Proc. Natl. Acad. Sci. USA} {\bf 84}, 7524--7528.
\bibitem{BW2} Bryngelson, J. D. \& Wolynes, P. G. (1989) {\it
J. Phys. Chem.} {\bf 93}, 6902--6915.
\bibitem{BW3} Bryngelson, J. D. \& Wolynes, P. G. (1990) {\it
Biopolymers} {\bf 30}, 177--188.
\bibitem{THI1} Honeycutt, J. D. \& Thirumalai, D. (1990) {\it
Proc. Natl. Acad. Sci. USA} {\bf 87}, 3526--3529.
\bibitem{SW} Sasai, M. \& Wolynes, P. G. (1992) {\it
Phys. Rev. A} {\bf 46}, 7979--7997.
\bibitem{Fuk} Fukugita, M., Lancaster, D. \& Mitchard, M. G. (1997) {\it 
 Biopolymers} {\bf 41}, 239--250.
\bibitem{BOSW} Bryngelson, J. D., Onuchic, J. N., Socci, N. D. \&
Wolynes, P. G. (1995) {\it Proteins} {\bf 21}, 167--195.
\bibitem{KS} Karplus, M. \& Sali, M. (1995) {\it
Curr. Opin. Struc. Biol.} {\bf 5}, 58--73.
\bibitem{DC} Dill, K. A. \& Chan, H. S. (1997) {\it Nature Structural
Biology} {\bf 4}, 10--19.
\bibitem{SHA4} Shakhnovitch, E. I. (1997) {\it Curr. Opin. Struc. Biol.}
 {\bf 7}, 29--40.
\bibitem{Brooks} Boczko, E. M. \& Brooks, C. L. III (1995) {\it Science} {\bf
269}, 393--396.
\bibitem{Vas} V{\'a}squez, M., N{\'e}methy, G. \& Scheraga, H. A. (1994)
{\it Chem. Rev.} {\bf 94}, 2183--2239.
\bibitem{MU} Berg, B. A. \& Neuhaus, T. (1991) {\it Phys. Lett.} {\bf
  B267}, 249--253.
\bibitem{ST} Lyubartsev, A. P., Martinovski, A. A., Shevkunov, S. V. \&
Vorontsov-Velyaminov, P. N. (1992) {\it J. Chem. Phys.} {\bf 96},
1776--1783.
\bibitem{ST2} Marinari, E. \& Parisi, G. (1992) {\it Europhys. Lett.} 
{\bf 19} 451--458.
\bibitem{HO} Hansmann, U. H. E. \& Okamoto, Y. (1993) {\it
J. Comp. Chem.} {\bf 14}, 1333--1338.
\bibitem{HO95a} Okamoto, Y. \& Hansmann, U. H. E. (1995) {\it
J. Phys. Chem.} {\bf 99}, 11276--11287.
\bibitem{HO96c} Hansmann, U. H. E., Okamoto, Y. \& Eisenmenger,
F. (1996) {\it Chem. Phys. Lett.} {\bf 259}, 321--330.
\bibitem{NNK} Nakajima, N., Nakamura, H. \& Kidera, A. (1997) {\it
J. Phys. Chem.} {\bf 101}, 817--824.
\bibitem{HO96b} Hansmann, U. H. E. \& Okamoto, Y. (1997)
{\it J. Comp. Chem.} {\bf 18}, 920--933.
\bibitem{H97a} Hansmann, U. H. E. (1997) ``Simulated Annealing with
 Tsallis Weights - A Numerical Comparison'', {\it Physica A}, in press.
\bibitem{HO96d} Hansmann, U. H. E. \& Okamoto, Y. (1997) 
 ``Generalized-Ensemble Monte Carlo Method for Systems with Rough Energy
 Landscape'', {\it Physical Review E} {\bf 56}, in press.
\bibitem{Tsa} Tsallis, C. (1988) {\it J. Stat. Phys.} {\bf 52}, 479--487.
\bibitem{FS} Ferrenberg, A. M. \& Swendsen, R. H. (1988) {\it Phys. Rev. 
Lett.} {\bf 61}, 2635--2638; (1989) {\it 
  Phys. Rev. Lett.} {\bf 63 }, 1658--1658(E), and references given in
  the erratum.
\bibitem{HO94_3} Hansmann, U. H. E. \& Okamoto, Y. (1994) {\it Physica
A} {\bf 212}, 415--437.
\bibitem{Molscript} Kraulis, P. J. (1991) {\it J. Appl. Cryst.} {\bf
24}, 946--950.
\bibitem{Raster3D} Bacon, D. \& Anderson, W. F. (1988) 
{\it J. Mol. Graphics}, {\bf 6}, 219--220;
Merritt, E. A. \& Murphy, M. E. P. (1994) {\it Acta Cryst.} 
{\bf D50}, 869--873.
\bibitem{EC3} Sippl, M. J., N{\'e}methy, G. \& Scheraga, H. A. (1984)
{\it J. Phys. Chem.} {\bf 88}, 6231--6233, and references therein.
\bibitem{KONF} Kawai, H., Okamoto, Y., Fukugita, M., Nakazawa, T. \&
  Kikuchi, T. (1991) {\it Chem. Lett.} {\bf 1991}, 213--216;
 Okamoto, Y., Fukugita, M., Nakazawa, T.\ \& Kawai, H.
 (1991) {\it Protein Engineering} {\bf 4}, 639--647.
\bibitem{MO} The program for calculation of solvent excluded volume was
 written by M.~Masuya and will be described in detail  elsewhere.
\bibitem{ELASS} Eisenhaber, F., Lijnzaad, P., Argos, P., Sander, C.
\& Scharf, M. (1995) {\it J. Comp. Chem.} {\bf 16}, 273--284.
\bibitem{Braun} Freyberg, B.~von \& Braun, W. (1991) {\it J.~Comp.Chem.}
                {\bf 12}, 1065--1076.
\bibitem{OK} Okamoto, Y., Kikuchi, T. \& Kawai, H.
  (1992) {\it Chem. Lett.} {\bf 1992}, 1275--1278.
\bibitem{EH96} Eisenmenger,~F. \& Hansmann,~U.H.E., (1997)
               {\it J.~Phys.~Chem.~B} {\bf 101}, 3304--3310.
\end{thebibliography}
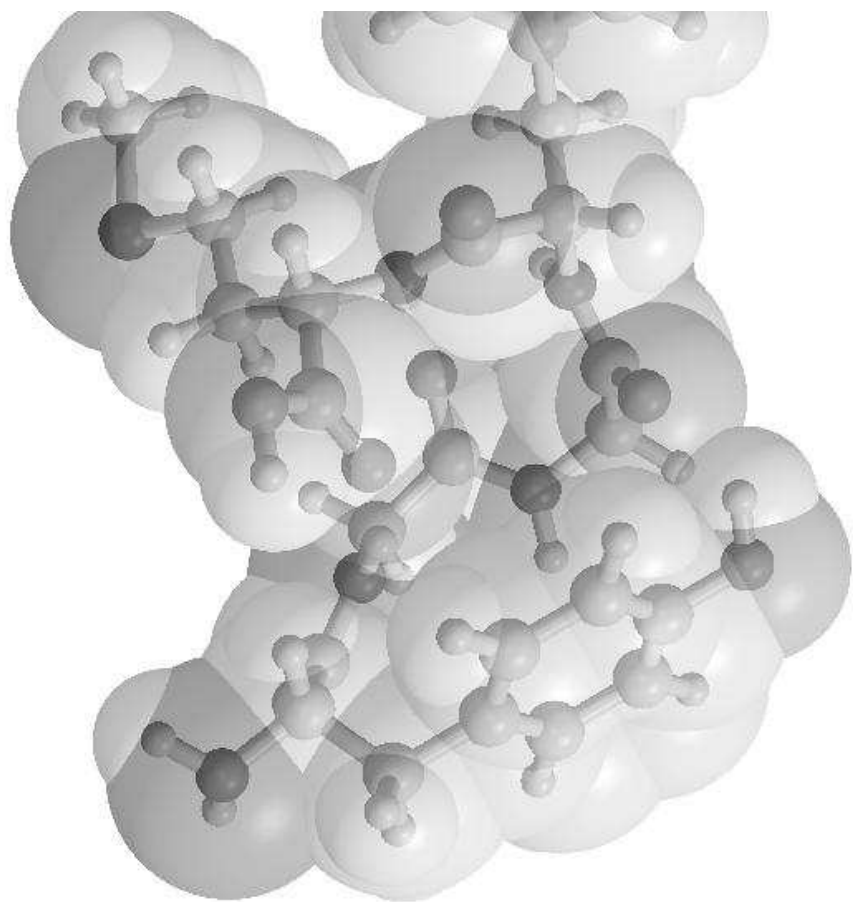
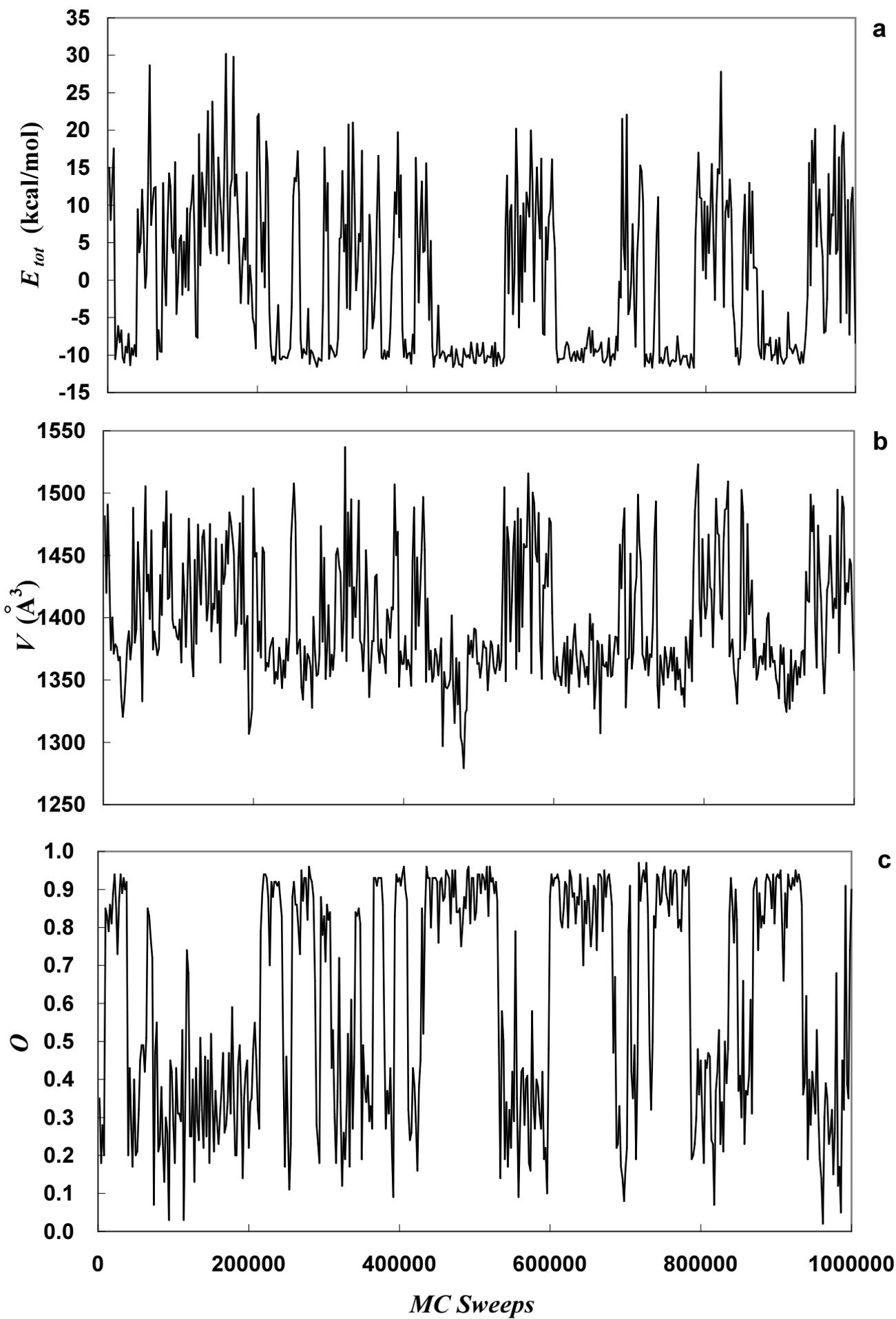
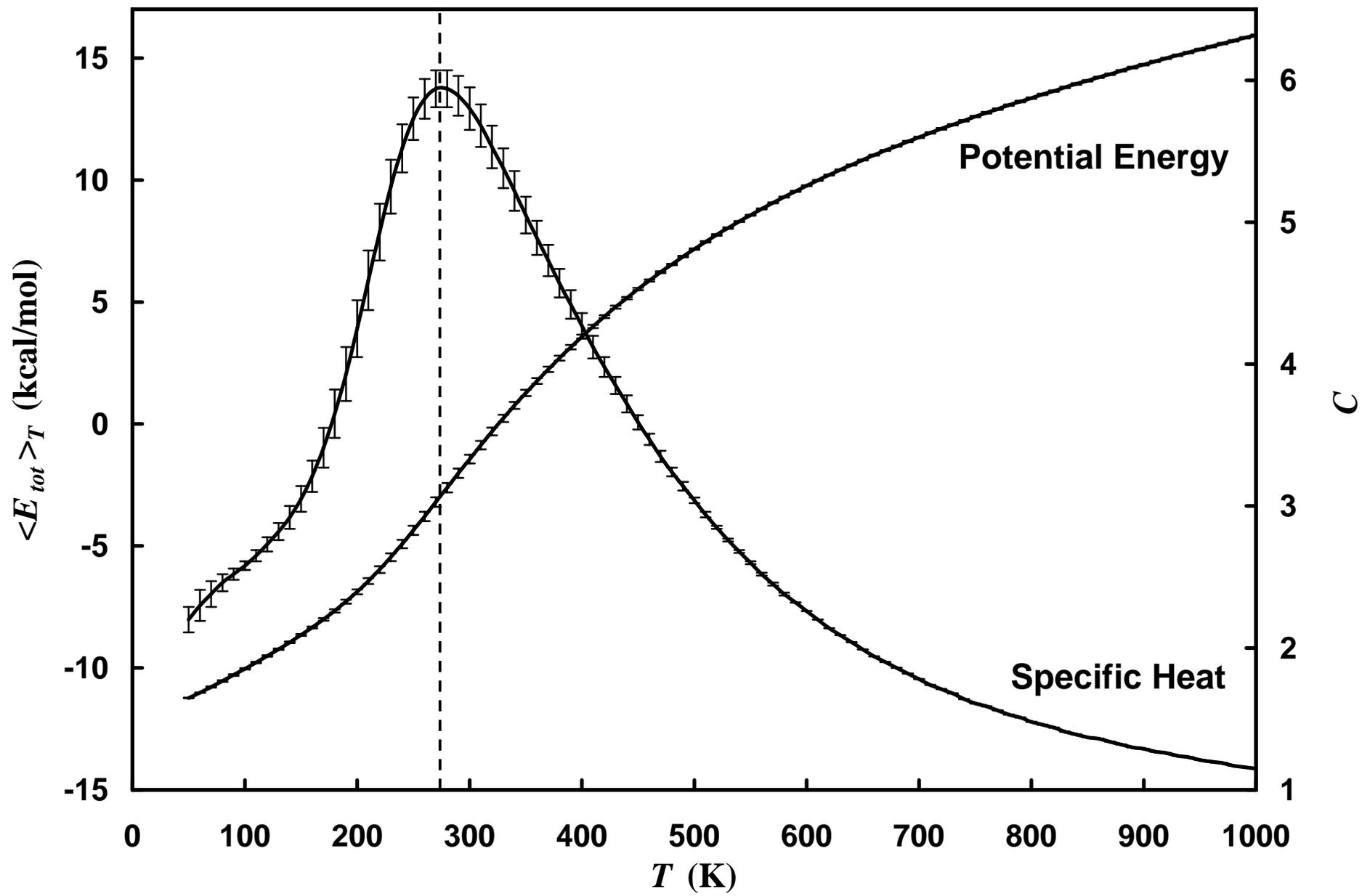
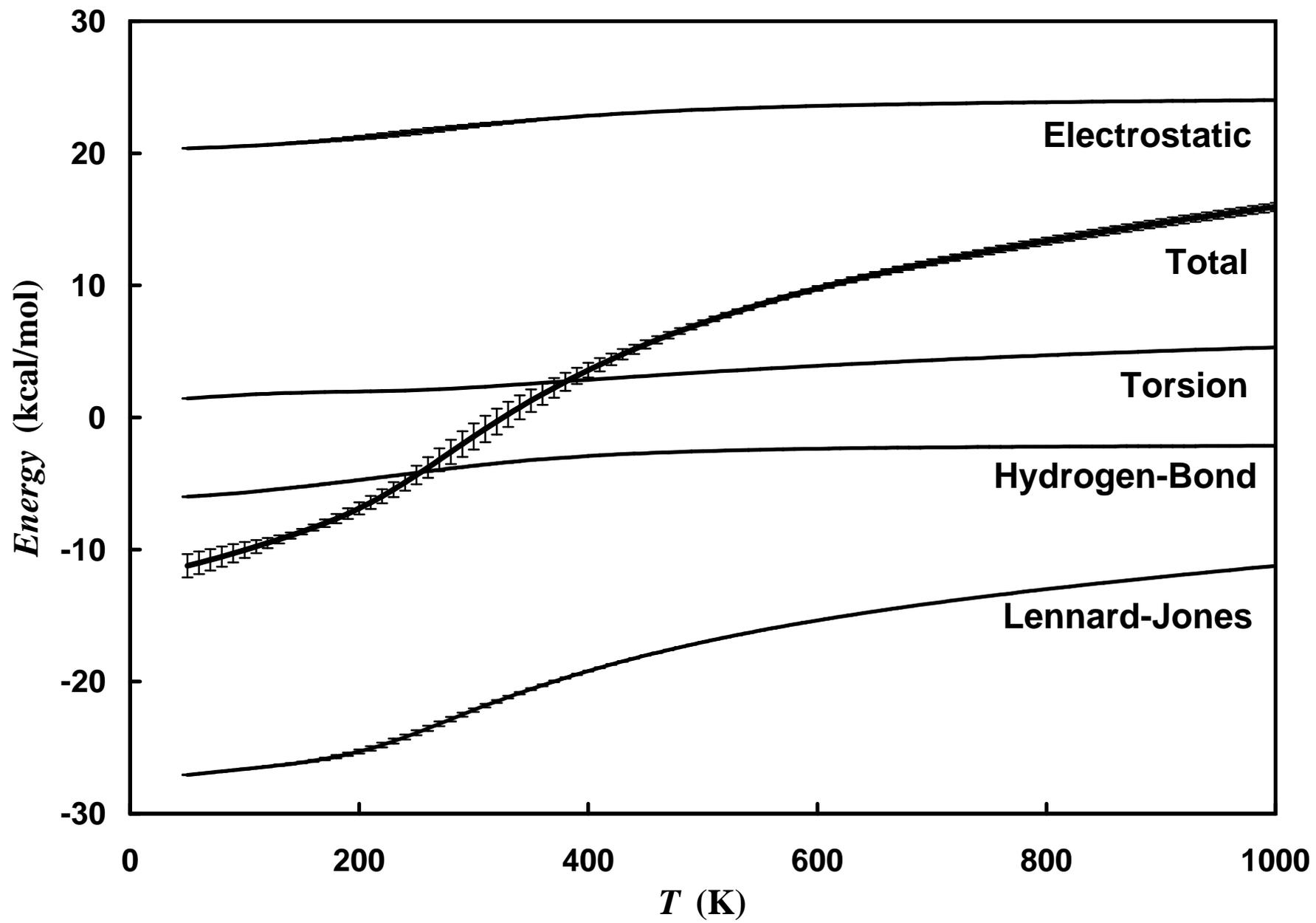
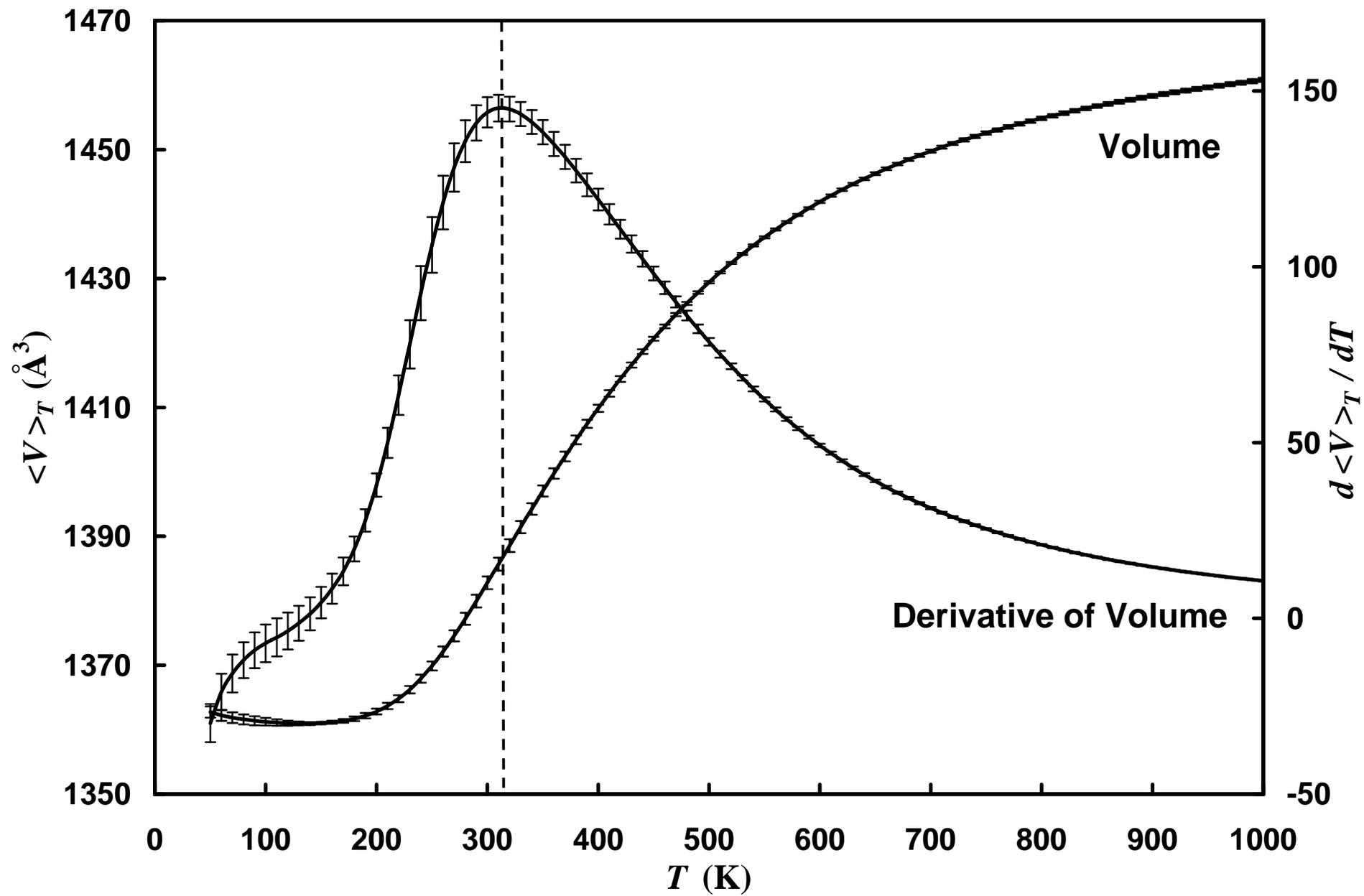
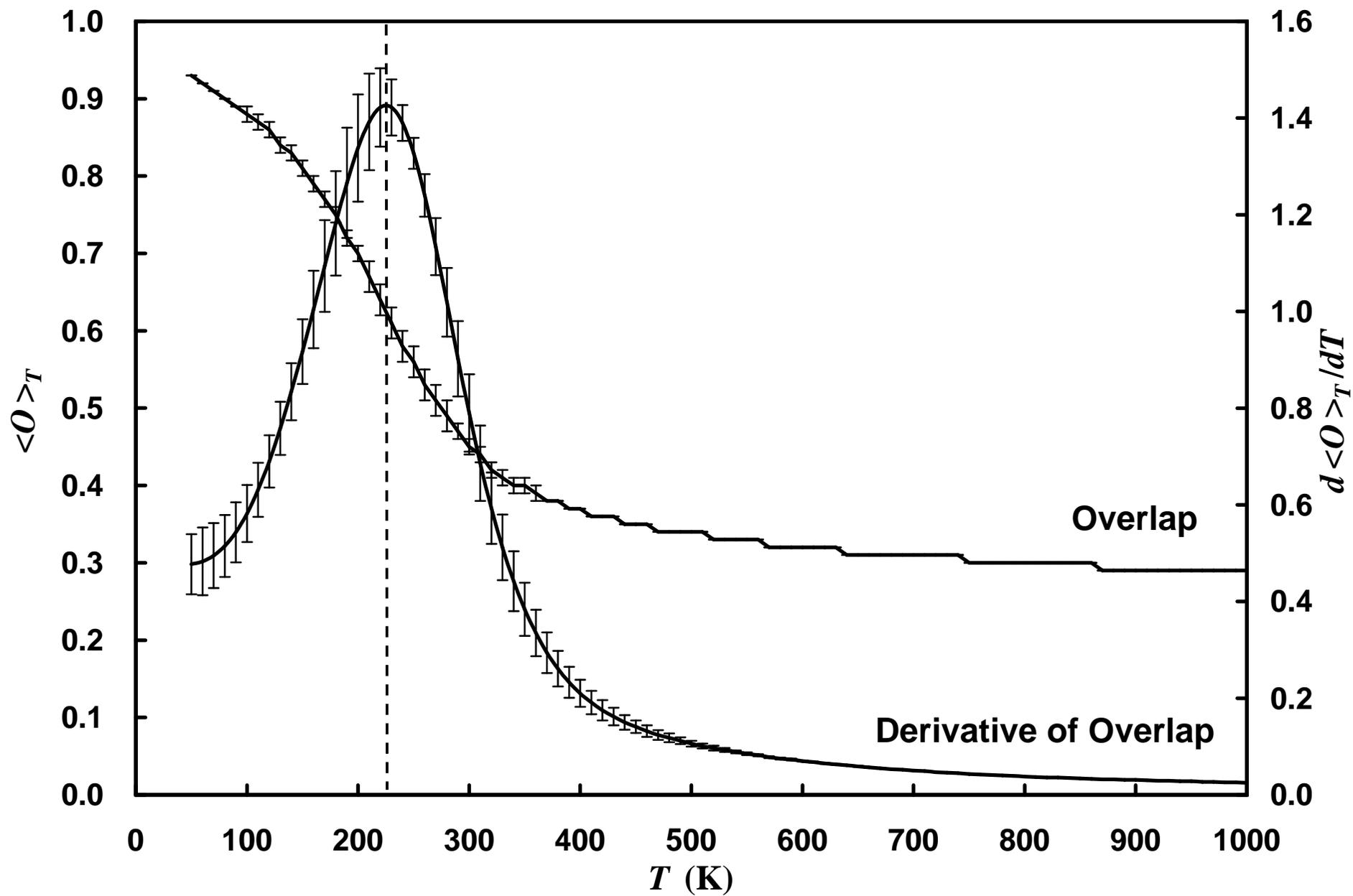
\end{document}